\documentclass[10pt,twocolumn,letterpaper]{article}

\usepackage{iccv}
\usepackage{times}
\usepackage{epsfig}
\usepackage{graphicx}
\usepackage{amsmath}
\usepackage{amssymb}

\usepackage[pagebackref=true,breaklinks=true,letterpaper=true,colorlinks,bookmarks=false]{hyperref}

\iccvfinalcopy 

\def\iccvPaperID{4141} 

\ificcvfinal\fi
\begin{document}

\title{Deep 3D-Zoom Net: Unsupervised Learning of Photo-Realistic 3D-Zoom}

\author{Juan Luis Gonzalez Bello and Munchurl Kim\\
Korea Advanced Institute of Science and Technology\\
{\tt\small juanluisgb@kaist.ac.kr mkimee@kaist.ac.kr}
\and
}

\maketitle

\begin{abstract}
The 3D-zoom operation is the positive translation of the camera in the Z-axis, perpendicular to the image plane. In contrast, the optical zoom changes the focal length and the digital zoom is used to enlarge a certain region of an image to the original image size. In this paper, we are the first to formulate an unsupervised 3D-zoom learning problem where images with an arbitrary zoom factor can be generated from a given single image. An unsupervised framework is convenient, as it is a challenging task to obtain a 3D-zoom dataset of natural scenes due to the need for special equipment to ensure camera movement is restricted to the Z-axis. In addition, the objects in the scenes should not move when being captured, which hinders the construction of a large dataset of outdoor scenes. We present a novel unsupervised framework to learn how to generate arbitrarily 3D-zoomed versions of a single image, not requiring a 3D-zoom ground truth, called the Deep 3D-Zoom Net. The Deep 3D-Zoom Net incorporates the following features: (i) transfer learning from a pre-trained disparity estimation network via a back re-projection reconstruction loss; (ii) a fully convolutional network architecture that models depth-image-based rendering (DIBR), taking into account high-frequency details without the need for estimating the intermediate disparity; and (iii) incorporating a discriminator network that acts as a no-reference penalty for unnaturally rendered areas. Even though there is no baseline to fairly compare our results, our method outperforms previous novel view synthesis research in terms of realistic appearance on large camera baselines. We performed extensive experiments to verify the effectiveness of our method on the KITTI and Cityscapes datasets.  \end{abstract}

\begin{figure}
  \centering
  \includegraphics[width=0.46\textwidth]{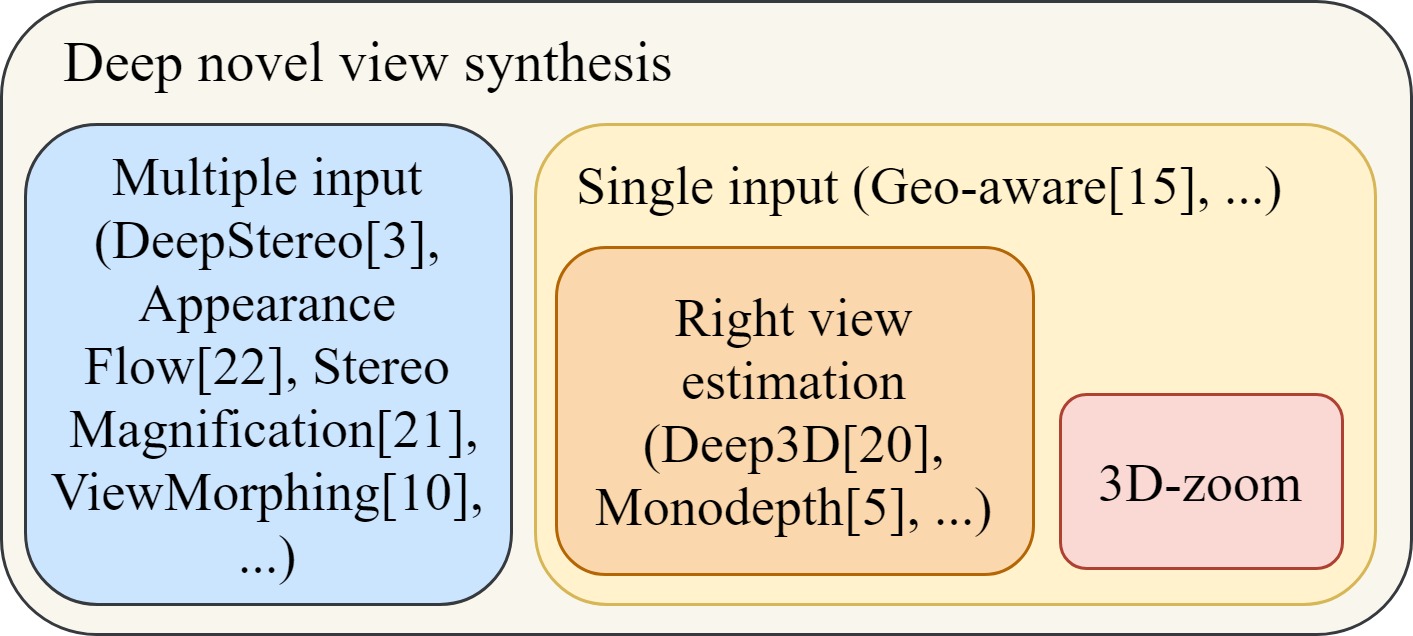}
  \caption{Categorization of Deep Novel View Synthesis. Our problem belongs to the novel view synthesis on a single image domain, and our pipeline is unsupervised.}
  \label{fig:venn}
\end{figure}

\begin{figure*}
  \centering
  \includegraphics[width=0.95\textwidth]{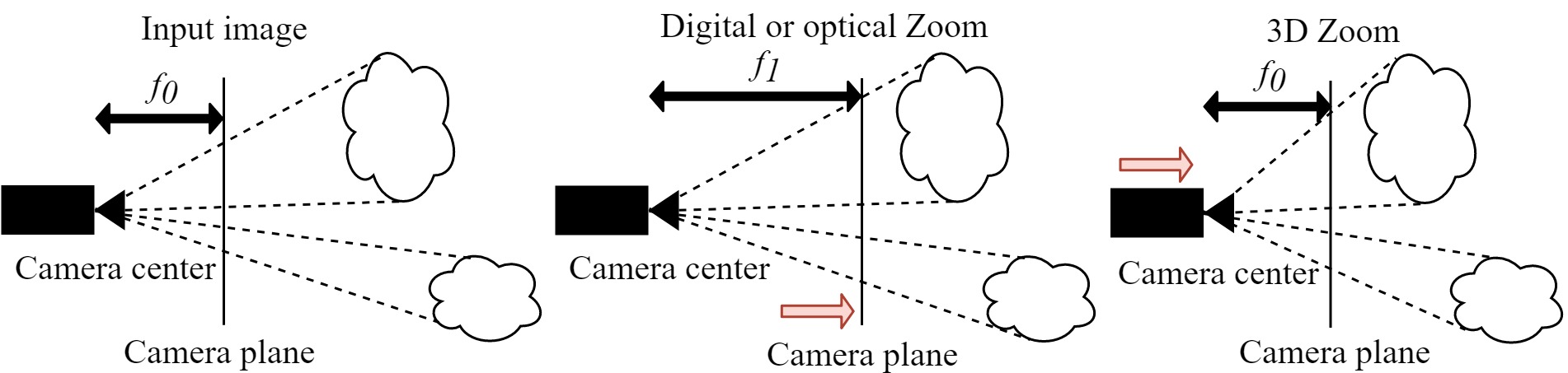}
  \caption{Optical Zoom vs 3D zoom}
  \label{fig:vs_zoom}
\end{figure*}

\section{Introduction}
Novel view synthesis is the task of hallucinating an image seen from a different camera pose given a single image or a set of input images. In natural images, this is a challenging task due to occlusions, ambiguities, and complex 3D structures in the scene. In addition, the larger the baseline (relative distance between input camera pose and target camera pose) the more challenging the problem becomes, as occlusions and ambiguities become dominant. New view synthesis finds applications in robotics, image navigation, augmented reality, virtual reality, cinematography, and image stabilization. 
There is a large body of literature that has studied the novel view synthesis problem for the multiple input image scenario, in both classical and learning based approaches. On the other hand, few works have tackled the problem of single input image novel view synthesis, which is a more complex task, as the deep understanding of the underlying 3D structure of the scene is needed to synthesize a new view. Finally, 3D-zoom is a subset of the novel view synthesis problem that has not been studied separately as exemplified in Figure \ref{fig:venn}.

3D-zoom is the positive translation of the camera in the Z-axis as depicted in Figure \ref{fig:vs_zoom}. In contrast, digital and optical zoom are close to a change in focal length and don't require any knowledge about the scene 3D geometry. 
Generating a 3D-zoom dataset with natural scene imagery is a challenging task. Special devices would need to be used to ensure translation is restricted to the Z-axis. In addition, moving objects would need to be masked or avoided as they would represent ambiguities for the 3d-zoom model. Alternatively, some available driving datasets could be used by filtering the sequences that move in a straight line. However, it does not guarantee camera pose changes to be restricted to the Z-axis neither the absence of moving objects between captures in the scene. For these reasons, we propose to learn 3D-zoom in an unsupervised fashion by utilizing a pre-trained disparity estimation network with transfer learning. Our 3D-Zoom Net is based on a fully convolutional network architecture that learns the under-laying 3D structure of the scene without the need of intermediate disparity as it is trained based on a novel back re-projection reconstruction cost that enforces both 3D geometry and natural appearance. Additionally, we include an adversarial network that acts as a {\it no-reference} measure that penalizes unnaturally rendered areas. Our proposed model, Deep 3D-Zoom Net, can perform inference of naturally looking 3D-zoomed images very fast. We show the efficacy of our proposed model in generating  3D-Zoomed images at various zoom factors on the KITTI \cite{kitti2012}\cite{kitti2015} and Cityscapes \cite{cityscapes} datasets. 

\section{Related works}
Novel view synthesis has been well studied over the years. We could define two types of algorithms, the multiple views, and the single view types. Multiple view algorithms are those that mainly rely on the correspondences between multiple input views to render the final synthetic view. In contrast, single image approaches rely on depth cues (textures, objects sizes, geometries, etc.) to model the 3D structure of the single image input and generate the novel view.

\subsection{Multiple input view synthesis}
\textbf{Classical approaches} for novel view synthesis rely on optimization techniques to render the new view. The method proposed by Chaurasia \etal \cite{localwarps} over-segments the input image into super-pixels to estimate depth via an optimization process. Super-pixels from multiple views are then warped (guided by the corresponding depth value) and blended to generate the novel view. In contrast, in \cite{content_warps}, Liu \etal, instead of estimating the depth map, used an off-the-shelf structure from motion algorithm to obtain the camera pose and fixed background points of a given video sequence in combination with traditional optimization techniques to directly estimate the warping operation for each input image. In \cite{onnewview}, Woodford \etal simultaneously solved for color and depth in the new view using a graph-cut-based optimizer for multi-label conditional random fields (CRFs). 

\textbf{Deep learning approaches.} Even though classical approaches succeed in their context, their performance is limited and proportional to the number of available input views. On the other hand, recent deep learning approaches have shown promising results for the novel view synthesis problem. The early work on natural real-world datasets of Flynn \etal \cite{deepstereo} takes multiple inputs and works on small patches to synthesize the target view. Their architecture, Deepstereo, divides the novel view synthesis problem into two branches, (1) selection volume and (2) image color estimation branches. The first performs image-based rendering (IBR) by learning how to blend the multiple input images. The second branch corrects the color for the target pixels. Their approach is very slow (taking up to seconds to perform inference). In the later work of Zhou \etal \cite{appflow}, based on the assumption that pixels in adjacent views are highly correlated, instead of estimating the view or the blending operation directly, they learned the warping operation to copy pixels from the input view into the new view. Their network is not fully convolutional and, whereas they showed good results on single object case, their model performs poorly for full scene synthesis. Similar to the classical approaches, the quality of the generated view in \cite{deepstereo} and \cite{appflow} is proportional to the number of input images. On the other hand, Ji \cite{deep_view_morphing} proposed Deep View Morphing, which receives two input images and estimates the intermediate view. This method first rectifies the pair, then estimates correspondences and visibility masks. These correspondences are used to warp the input images into the intermediate pose and the visibility masks are used to blend them together. This work resembles the video frame interpolation work by \cite{super_slow_mo}. Similarly, the model proposed by Zhou \etal \cite{stereo_mag} takes two images as input and generates new views along and beyond the baseline, and in a similar way to \cite{deepstereo, deep3d, geo_aware} a multi-channel representation of the input image is learned, but instead of being a selection volume, it is a multiplane image with corresponding alpha channels. This multiplane image can then be used to synthesize multiple new views by applying planar transformations.

\subsection{Single input view synthesis}
\textbf{Classical approaches} for single input view synthesis have shown very limited performance under several assumptions. In \cite{tourin}, Horry \etal used depth priors from user input to model the scene 3D information. Hoiem \etal proposed Photopop-up \cite{photopop}, which aims to statistically model geometric classes defined by the scene’s objects' orientations. By coarse labeling, they achieve decent performance on large structures like landscapes or buildings but seriously fail on estimating the 3D structure of thin and complex objects. The more recent work of Rematas \etal \cite{Rematas} takes a single image object and a 3D model prior. Their model learns how to align the 3D model with the input view and estimates each output pixel in the novel view as a linear combination of the input pixels. Performance is far from real-time and limited to the 3D models of single objects in the collection.

\textbf{Deep learning approaches.} Single image novel view synthesis has been greatly benefited by deep learning approaches. The recent work of Liu \etal \cite{geo_aware} tried to solve the problem by incorporating four networks for the disparity, normals, selection volume estimation, and image refinement, respectively. The predicted disparities and normals are combined with a super-pixel over-segmentation mask like in \cite{localwarps} to create a fixed number of homographies which produce warped images from the monocular input. These images are blended together, weighted by the estimated selection volume, which is also pre-warped by the corresponding homographies. The disparity and normals network follow the UNET architecture, whereas the selection volume is estimated from the up-scaling of deep features from an encoder-like architecture, similar to \cite{deepstereo}. In addition, the refinement network further improves the final result. In a subclass of novel view generation algorithms, Deep3D \cite{deep3d} reduces the scope of novel view synthesis to estimate the corresponding right view from a single left input image. Similar to \cite{deepstereo, geo_aware}, Deep3d produces a probabilistic disparity map to blend multiple left and right shifted versions of the input left view to generate a fixed synthetic right view. Deep3D limits itself to produce low-resolution probabilistic disparity maps due to its non-fully convolutional architecture. By enforcing geometry constraints, CNNs can be trained to learn disparity in an unsupervised fashion from stereo inputs by minimizing a reconstruction loss between a synthesized new view and the input view. Godard \etal introduced a monocular disparity estimation network, the Monodepth \cite{godard}, where their left-right consistency loss term greatly improved performance. However, their network could not estimate a complete disparity map in a single pass. Gonzalez and Kim \cite{gonzalez} further improved the performance of unsupervised disparity learning architectures by modeling ambiguities in the disparity maps and enabling full disparity estimations in a single pass, even with almost one-third of numbers of parameters in comparison with \cite{godard}. We make use of their pre-trained models to train our 3D-zoom architectures.

\textbf{3D-zoom: Unsupervised single image close-up view synthesis}. 3D-zoom is a subset of the single image novel view synthesis, where the camera pose change is restricted to be in the Z-axis only. Our novel work is the first to isolate and solve the 3D-zoom learning problem in an unsupervised fashion. We are able to learn novel view synthesis by modeling 3D-zoom as a blending operation between multiple up-scaled versions of a single input image. Our novel back re-projection reconstruction loss facilitates learning the under-laying 3D structure of the scene while preserving the natural appearance of the generated 3D-zoomed view, even while performing very fast inference.

\begin{figure*}
  \centering
  \includegraphics[width=\textwidth]{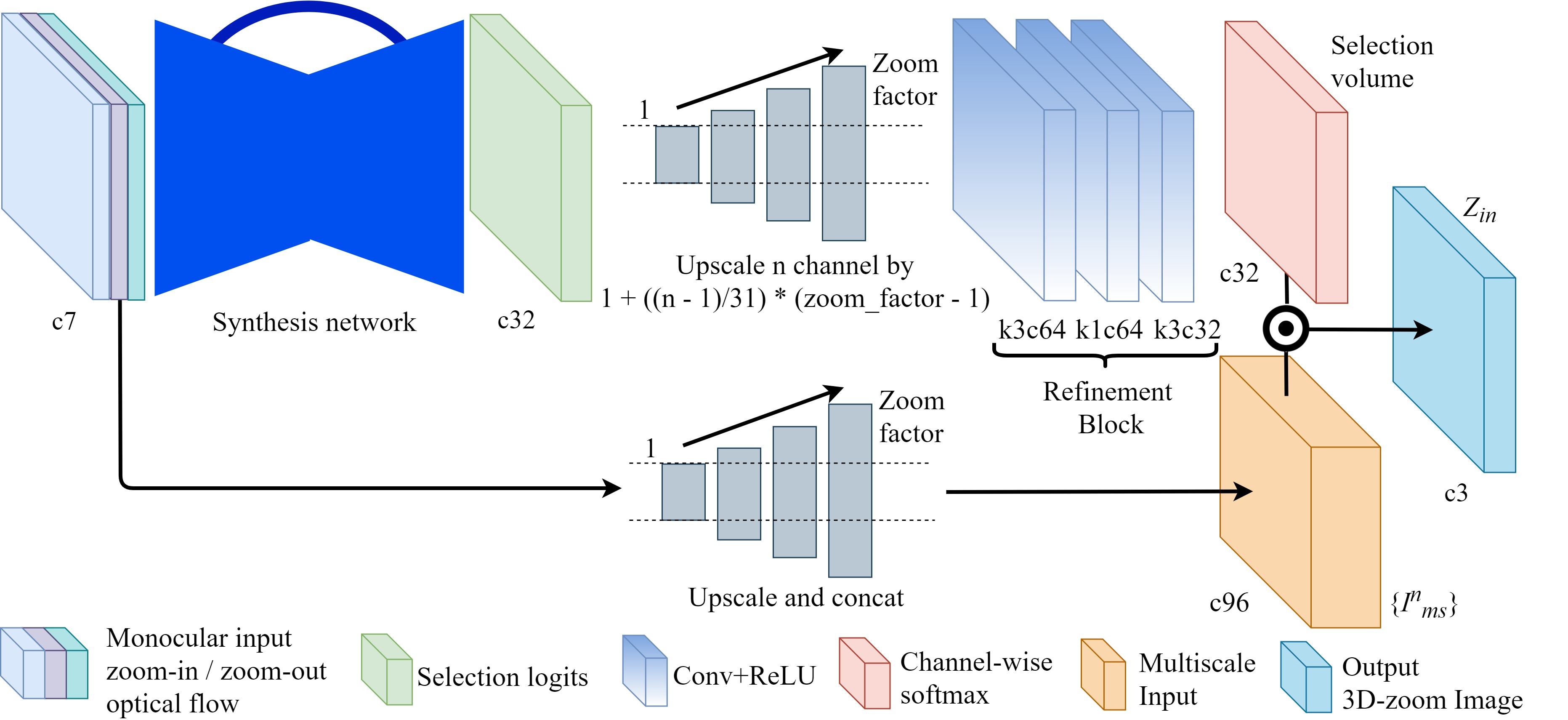}
  \caption{Deep 3D-Zoom Net for inference. It consists of synthesis network, the refinement block and the blending operation.}
  \label{fig:model}
\end{figure*}

\subsection{3D-zoom}
3D-zoom can be defined as the positive translation of the camera in the Z-axis. From the pinhole camera model, the following basic trigonometric relationship can be obtained
\begin{equation} \label{eq:prop_t}
\tan{\theta} = \dfrac{x_c}{f} = \dfrac{X_w}{Z_w}
\setlength{\abovedisplayskip}{4pt}
\setlength{\belowdisplayskip}{4pt}
\end{equation}
where $\theta$ is the angle measured from the principal axis to the camera plane coordinate, $x_c$ is the $x$ component of the camera plane coordinate, $X_w$ is $x$ component of the world coordinate, $f$ is the focal length, and $Z_w$ is the Z-axes component of the world coordinate. The projection in the camera plane can be defined as
\begin{equation} \label{eq:prop_t1}
x_c = \dfrac{X_wf}{Z_w}
\setlength{\abovedisplayskip}{4pt}
\setlength{\belowdisplayskip}{4pt}
\end{equation}
where $Z_w$ or “depth” is inversely proportional to disparity “$D$” and directly proportional to the focal length $f$ and the separation between stereo cameras $s$, and is defined as
\begin{equation} \label{eq:z_w}
Z_w = \dfrac{sf}{D}
\setlength{\abovedisplayskip}{4pt}
\setlength{\belowdisplayskip}{4pt}
\end{equation}
Therefore, the projection in the camera plane $x_c$ can be re-written as
\begin{equation} \label{eq:x_c0}
x_c = \dfrac{X_wD}{s}
\setlength{\abovedisplayskip}{4pt}
\setlength{\belowdisplayskip}{4pt}
\end{equation}
We can generalize the projection for any camera setup by taking the proportionality and furthermore by using a normalized disparity map $Dn$. This is defined as
\begin{equation} \label{eq:x_c1}
x_c \propto X_wDn
\setlength{\abovedisplayskip}{4pt}
\setlength{\belowdisplayskip}{4pt}
\end{equation}
Finally, any change in world coordinates $\Delta X_w$  (e.g. 3D-zoom) is projected into the camera plane weighted by the normalized disparity map as
\begin{equation} \label{eq:x_c2}
\Delta x_c \propto \Delta X_w Dn
\setlength{\abovedisplayskip}{4pt}
\setlength{\belowdisplayskip}{4pt}
\end{equation}
This allows us to use the normalized disparity map to weight the zoom-in optical flow, which is a critical step in our novel back re-projection reconstruction loss function. In other words, up-scaling of objects/pixels in 3D-zoom is linearly proportional to their disparity values. If an object is closer to the camera, it will have a larger disparity value, thus, leading to high up-scaling. Similarly, a faraway object from the camera will have a low disparity, leading to small or no up-scaling.

\section{Method}
As demonstrated in the previous section, 3D-zoom can be understood as a 3D-geometry-dependant up-scaling operation. Therefore, we model the synthesis problem as learning the blending operation between multiple up-scaled versions of the single input image $I_{ms}$. The blending operation consists of an element-wise multiplication, denoted by $\odot$, between the $n$-th channel of the selection volume $Selection\_vol^n(\cdot)$ and $I^n_{ms}$, followed by a summation along the channel axis, defined as 
\begin{equation} \label{eq:z_in}
Z_{in} = \sum_{n=1}^{N}I^n_{ms} \odot Selection\_vol^n(I, f_{in}, f_{out})
\end{equation}
where $Z_{in}$ is the output 3D-zoomed image, $I$ is the single image input, $f_{in}$ is the uniform zoom-in optical flow, $f_{out}$ is the uniform zoom-out optical flow, $N$ is the number of channels of the selection volume, and $I_{ms}$ represents the multiple bilinear up-scaled versions of the input image from unity ($upscale\_ratio=1$) to the target zoom factor ($upscale\_ratio=zoom\_factor$). $I_{ms}$, $f_{in}$ and $f_{out}$ are defined in Equations \ref{eq:i_ms}, \ref{eq:f_in} and \ref{eq:f_out} respectively. 
\begin{equation} \label{eq:i_ms}
I^n_{ms} = upscale(I, 1 + \dfrac{n}{N}(zoom\_factor-1))
\end{equation}
\begin{equation} \label{eq:f_in}
f_{in} = (1 - zoom\_factor)i\_grid
\end{equation}
\begin{equation} \label{eq:f_out}
f_{out} = (1/zoom\_factor - 1)i\_grid
\end{equation}
where $i\_grid$ is a uniform grid defined by $i\_grid_{ij} = (i, j)$.

\begin{figure*}
  \centering
  \includegraphics[width=0.9\textwidth]{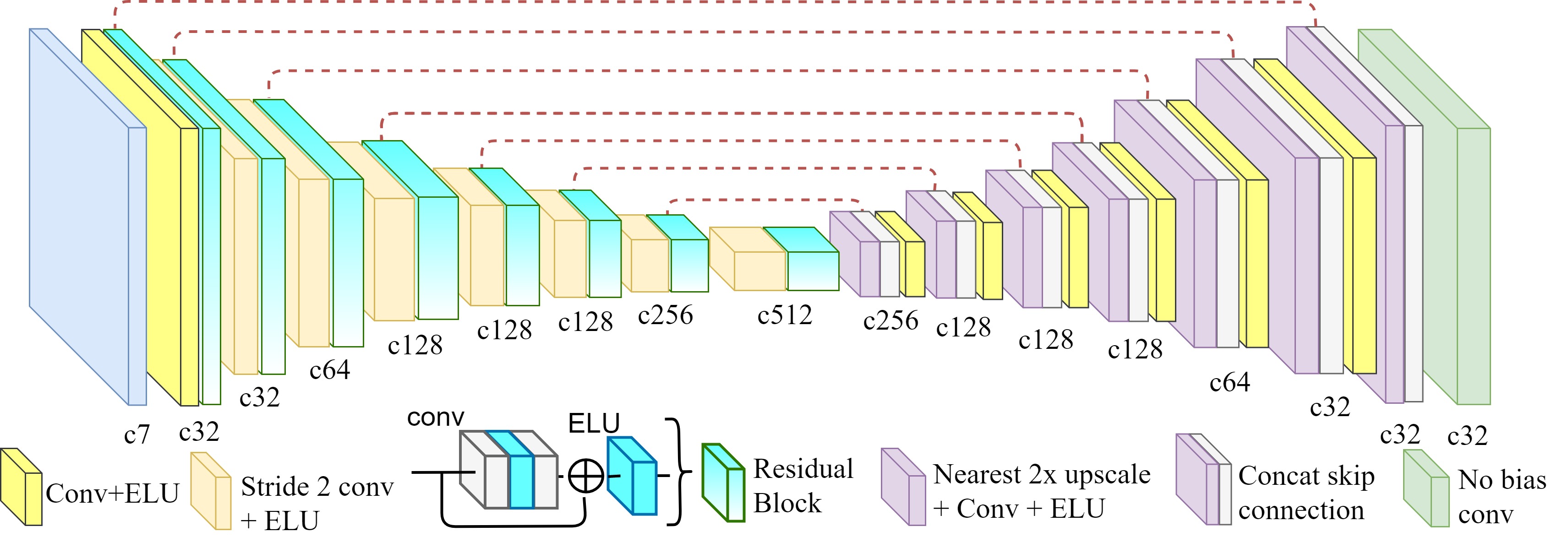}
  \caption{Fully convolutional synthesis network. Our synthesis network follows the UNET architecture with residual blocks.}
  \label{fig:synet}
\end{figure*}

\subsection{Network architecture - Deep 3D-Zoom Net}
Our proposed network architecture, which we call $Deep 3D-Zoom Net$, is shown in Figure \ref{fig:model} and is composed by an auto-encoder synthesis network, a refinement block, and the final blending operation. Our architecture takes a single image $I$, along with the uniform zoom-in and zoom-out optical flows $f_{in}$ and $f_{out}$ as a concatenated input. The synthesis network extracts the under-laying 3D-structure from the single image and generates the selection logits, which are the precursors of the selection volume. The selection logits are then bi-linearly expanded in a similar way to \{$I^n_{ms}$\} and fed into the refinement block which models the local relationships between the channels of the selection logits after being expanded. Finally, a channel-wise softmax is applied to generate the final selection volume. The selection volume is used to blend the multi-scale inputs \{$I^n_{ms}$\} into the final 3D-zoomed-in image $Z_{in}$ as described in Equation (\ref{eq:z_in}). In contrast with \cite{deepstereo} and \cite{deep3d} we apply the expansion operation to the selection logits instead of directly applying softmax on them. Also, in contrast with \cite{geo_aware}, our refinement block works on the selection logits instead of the synthetic image. Modeling the local relationships of the blending volume is essential under the absence of the 3D-zoomed ground truth. In contrast with other novel view synthesis techniques like \cite{deepstereo, deep3d, deep_view_morphing}, which estimate a fixed novel view depending on the input views, our network architecture allows for novel view generation with arbitrary zoom factors. 

\begin{figure}
  \centering
  \includegraphics[width=0.48\textwidth]{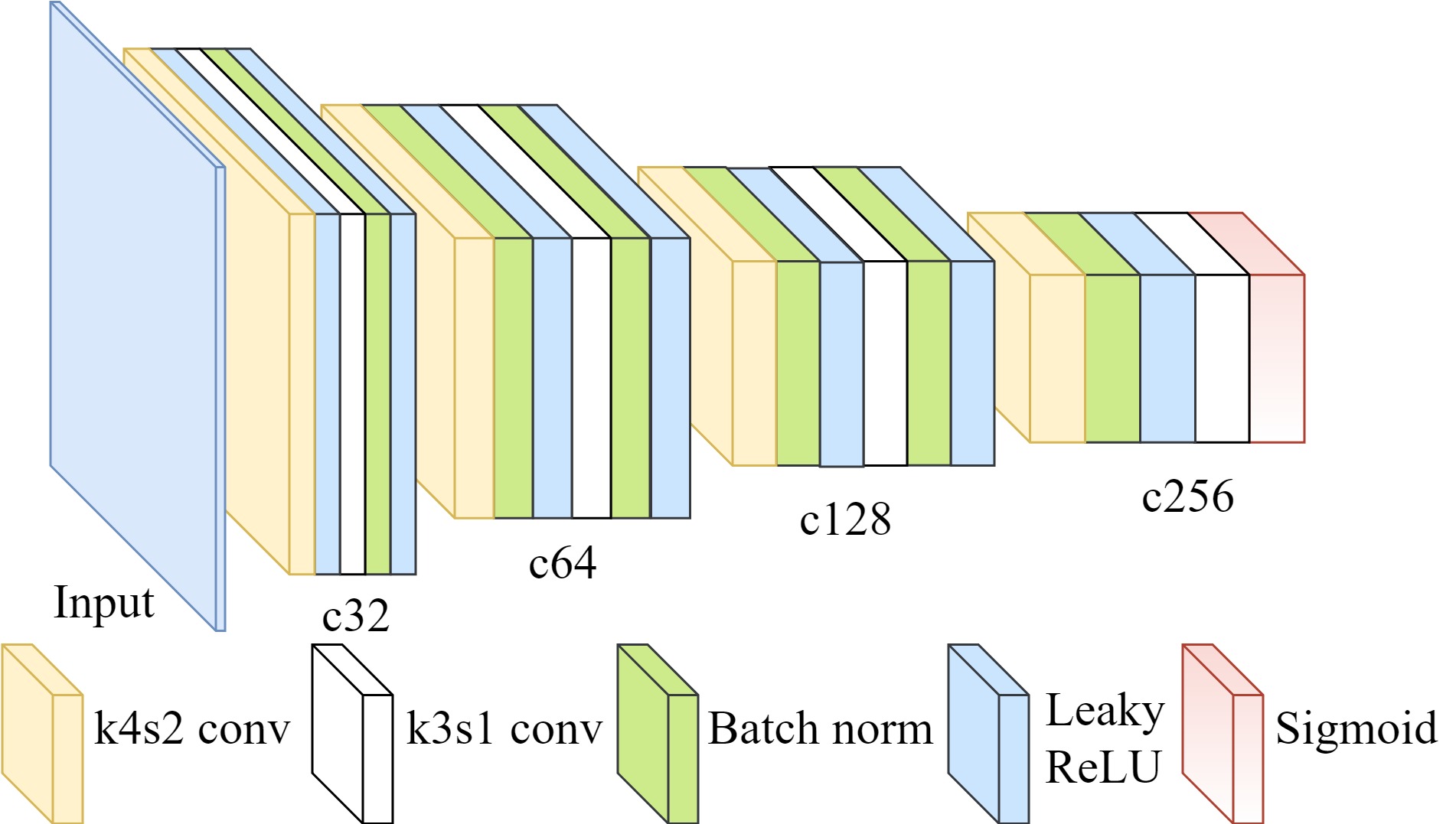}
  \caption{Our fully convolutional patch discriminator network.}
  \label{fig:disc}
\end{figure}

\subsubsection{Synthesis network}
A UNET-like architecture is used to extract the underlying 3D structure from the single image input. We designed the encoder part of our synthesis network inspired by the light encoder architecture of Gonzalez and Kim \cite{gonzalez}, which contains only 3x3 convolutions and residual blocks. Our fully convolutional synthesis network is depicted in Figure \ref{fig:synet}. Our synthesis network is fed with the channel-wise concatenated single input view and optical flows. Strided convolutions followed by residual blocks are used to downscale and extract relevant features trough seven stages. The decoder part of our synthesis network combines local and global information by adopting skip connections from the encoder part and performing nearest up-scaling plus 3x3conv and exponential linear unit (ELU) until the target resolution is achieved. Note that our fully convolutional network allows for high-resolution selection volumes, in contrast with \cite{deepstereo, deep3d, appflow}, where their fully connected layers fix the size of the input patch. The output of our synthesis network constitutes the $N$ channels of selection logits. We set $N=32$ for all our experiments.

\begin{figure*}
  \centering
  \includegraphics[width=\textwidth]{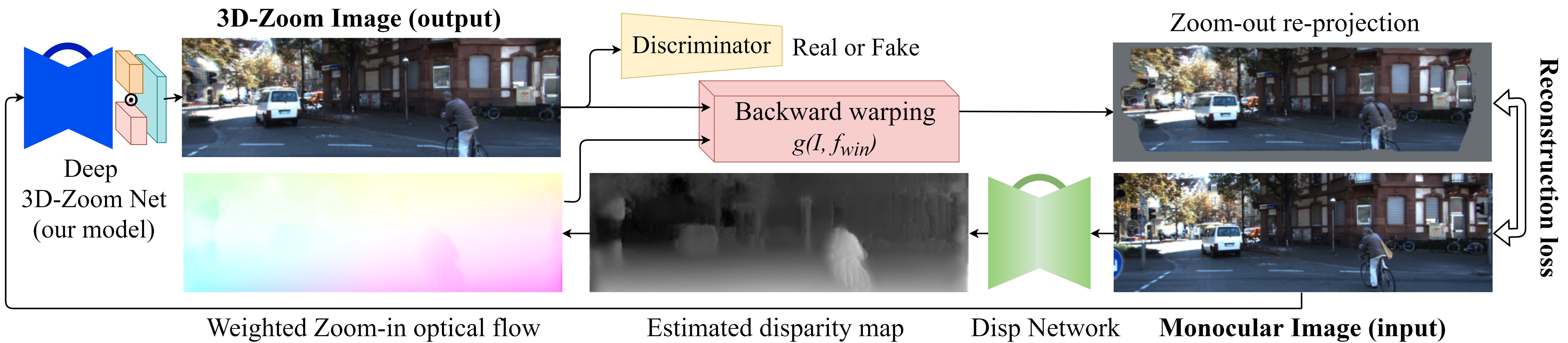}
  \caption{Training strategy for Deep 3D-Zoom Net. The re-projection reconstruction loss is computed between the zoomed-out re-projection and the input image. An adversarial loss is computed over the network output.}
  \label{fig:loss_diagram}
\end{figure*}

\subsection{Training strategy}
Due to the unsupervised nature of our problem, we have adopted a transfer learning strategy that relies on a novel back re-projection reconstruction loss, that allows the network not only to learn the underlying 3D structure but also to maintain a natural appearance. Figure \ref{fig:loss_diagram} depicts our training strategy. Given a single input image, a pre-trained disparity estimation network is used to estimate monocular disparity during training, which, once normalized, can be used to generate a weighted zoom-in optical flow by element-wise multiplication with the uniform zoom-in optical flow $f_{in}$, as defined in Equation (\ref{eq:3d_fin}) and depicted in Figure \ref{fig:loss_diagram}. We feed our network with the same monocular input image and estimate a 3D-zoomed version $Z_{in}$. By back re-projecting the estimated $Z_{in}$ image into the input image via a backward warping operation $g(\cdot)$, we obtain a zoomed-out image $Z_{out}$, defined in Equation (\ref{eq:z_out}), that can be compared against the input image. The resulting error can then be minimized to end-to-end train our model.
\begin{equation} \label{eq:z_out}
Z_{out}=g(Z_{in}, f_{win})
\end{equation}
\begin{equation} \label{eq:3d_fin}
f_{win}= f_{in} \odot Dn
\end{equation}
where $Dn = disp\_network(I) / max(disp\_network(I))$, and $disp\_network(\cdot)$ is the output of the disparity network from \cite{gonzalez}.
As depicted in Figure \ref{fig:loss_diagram}, the $g(\cdot)$ is open not capable of reconstructing the image borders $Z_{out}$. We define a dis-occlusion mask that takes this into account and lets the loss function to ignore those areas in cost calculations. The dis-occlusion mask is defined as
\begin{equation} \label{eq:disocc_mask}
disocc\_mask_{ij} =
\left \{
  \begin{tabular}{cc}
  $0$ & $if\ i + f_{inij} \odot Dn_{ij} > W$ \\
  $0$ & $if\ i + f_{inij} \odot Dn_{ij} < 0$ \\
  $0$ & $if\ j + f_{inij} \odot Dn_{ij} > H$ \\
  $0$ & $if\ j + f_{inij} \odot Dn_{ij} < 0$ \\
  $1$ & $o.w.$ \\
  \end{tabular}
\right.
\end{equation}
where $H$ and $W$ are the input image height and width respectively. Applying the dis-occlusion mask we get the complete zoom-out image $\Tilde{Z}_{out}$, which is given by
\begin{equation} \label{eq:z_out_p}
\Tilde{Z}_{out}=disocc\_mask \odot Z_{out} + (1 - disocc\_mask) \odot I
\end{equation}

\subsubsection{Reconstruction loss}
Our reconstruction loss is defined as a combination of two terms, appearance loss and perceptual loss, as
\begin{equation} \label{eq:rec_loss}
l_{rec} = 0.8l_{ap} + 0.2l_{p}
\end{equation}
\textbf{Appearance loss}. The appearance loss enforces the image $\Tilde{Z}_{out}$ to be similar to the input image $I$, and can be defined by the weighted sum of the $l_1$ and $ssim$ loss terms (a weight of $\alpha = 0.85$ was used) as
\begin{equation} \label{eq:ap_loss}
l_{ap} = \alpha||I - \Tilde{Z}_{out}||_1 + (1 - \alpha)SSIM(I, \Tilde{Z}_{out})
\end{equation}
\textbf{Perceptual loss}. Perceptual loss \cite{perceptual} is ideal to penalize deformations, textures and lack of sharpness. Three layers, denoted as $\phi^l$, from the pre-trained $VGG19$ \cite{vgg} ($relu1\_2, relu2\_2, relu3\_4$) were used as follows:
\begin{equation} \label{eq:vgg_loss}
l_p = \sum_{l=1}^{3} ||\phi^l(I)-\phi^l(\Tilde{Z}_{out})||_1
\end{equation}

\subsubsection{Adversarial loss}
In addition to not counting on a 3D-zoomed ground truth (GT), the disparity map, needed for training only, is not perfect as it is obtained from a pre-trained network. To mitigate this issue, we incorporate a discriminator network that acts as a {\it no-reference} penalty function for unnaturally rendered areas. Our discriminator network is depicted in Figure \ref{fig:disc}. It consists of four stages of strided Conv-BN-LReLU-Conv-BN-LReLU (BN: batch norm, LReLU: leaky relu) through which the single image input is down-scaled from 256x256 to 16x16, where the final activation function is not leaky ReLU but sigmoid. Since the 3D-zoom ground truth is not available, our networks cannot be trained on the recent WGANGP \cite{wgangp} configuration, as the gradient penalty term in it could not be estimated. Instead, the traditional patch-GAN training technique was used with the mean square error (MSE) loss. Our novel back re-projection reconstruction loss with an adversarial loss is defined as 
\begin{equation} \label{eq:rec_loss_gan}
l_{rec} = 0.8l_{ap} + 0.2l_{p} + 0.02l_d
\end{equation}
where $l_d$ is the adversarial loss, $l_{ap}$ is the appearance loss, and $l_p$ is the perceptual loss. 
While the generator network, Deep 3D-Zoom Net, is trained to minimize the probability of the generated image to be classified as fake, the discriminator is trained to correctly classify real and fake images. This can be formulated as minimizing
\begin{equation} \label{eq:d_loss}
l_D = mse(D(Z_{in}), \mathbf{0}) + mse(D(I), \mathbf{1})
\end{equation}
where $D$ indicates the discriminator network and $l_D$ indicates the discriminator loss. The real images are sampled from the inputs to the Deep 3D-Zoom Net, and the fake images sampled from the Deep 3D-Zoom Net outputs.

\begin{figure*}
  \centering
  \includegraphics[width=\textwidth]{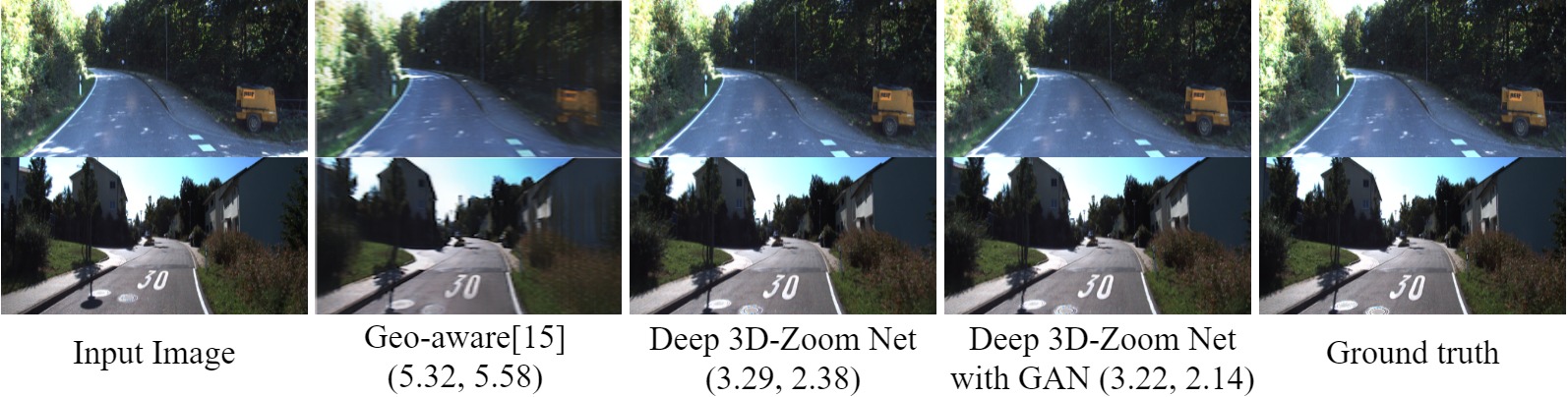}
  \caption{Results on KITTI2012 / NIQE for sampled images (top, bottom) and subjective comparison with the results showed in \cite{geo_aware} for visual quality only. In terms of natural image generation, our Deep 3D-Zoom Net outperforms geometric-aware networks with no visible artifacts for the equivalent zoom factors (1.6 top, 2.4 bottom). Note ground truth is just for reference, and was not used to train our model.}
  \label{fig:vs_geo}
\end{figure*}

\section{Results}
We perform extensive experiments to verify the effectiveness of our proposed model and training strategy on the KITTI2015 \cite{kitti2015} dataset which contains 200 binocular frames and sparse disparity ground truth obtained from velodyne laser scanners and CAD models. An ablation study is performed by training and testing our networks with and without the refinement block, perceptual loss, and adversarial loss to prove the efficacy of each of them. Additionally, we test our Deep 3D-Zoom Net on the Cityscapes \cite{cityscapes} dataset, a higher resolution urban scene dataset, to demonstrate it can generalize to previously unseen scenes.

\begin{figure*}
  \centering
  \includegraphics[width=\textwidth]{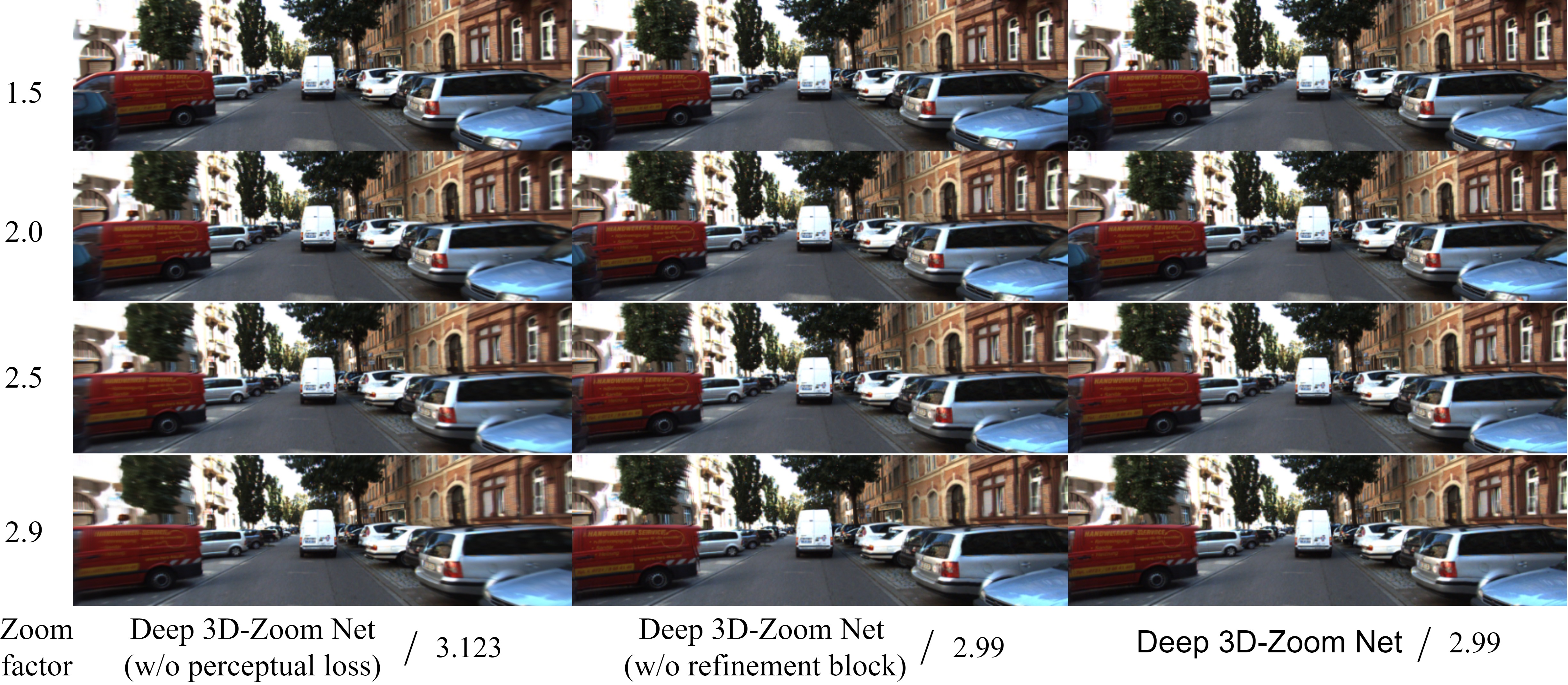}
  \caption{Results from ablation studies / NIQE score. A progressive improvement in terms of structure and sharpness can be appreciated from our model trained without perceptual loss to our model trained with perceptual loss and refinement block.}
  \label{fig:ablation}
\end{figure*}

\subsection{Implementation details}
We used the Adam \cite{adam} optimizer with the recommended betas for image synthesis (beta1 = 0.5 and beta2 = 0.9). Our models were trained for 50 epochs with an initial learning rate of 0.0001 for the generator, and 0.00001 for the discriminator. The mini-batch size was set to 8 images. The learning rate was halved at epochs 30 and 40. The following data augmentations on-the-fly were performed: random crop (256x256), random horizontal flips, random gamma, brightness and color shifts. All models were trained on the KITTI split \cite{godard}, which consists of 29,000 stereo pairs spanning 33 scenes from the KITTI2012 dataset \cite{kitti2012}. As can be seen in Figure \ref{fig:loss_diagram}, the dis-occlusion area grows along with the zoom factor, and this limits the effective area to train the network. Therefore, to properly train the network on higher zoom factors, we need to train the model on large zoom factors more often than small zoom factors. To achieve this, the zoom factor for each image in the mini-batch is randomly sampled from a left-folded normal distribution with $\mu=max\_zoom\_factor$ and $\sigma = 1$ to ensure larger zoom factors are trained more often. We set the $max\_zoom\_factor=3$ for all our experiments.

\begin{figure*}
  \centering
  \includegraphics[width=\textwidth]{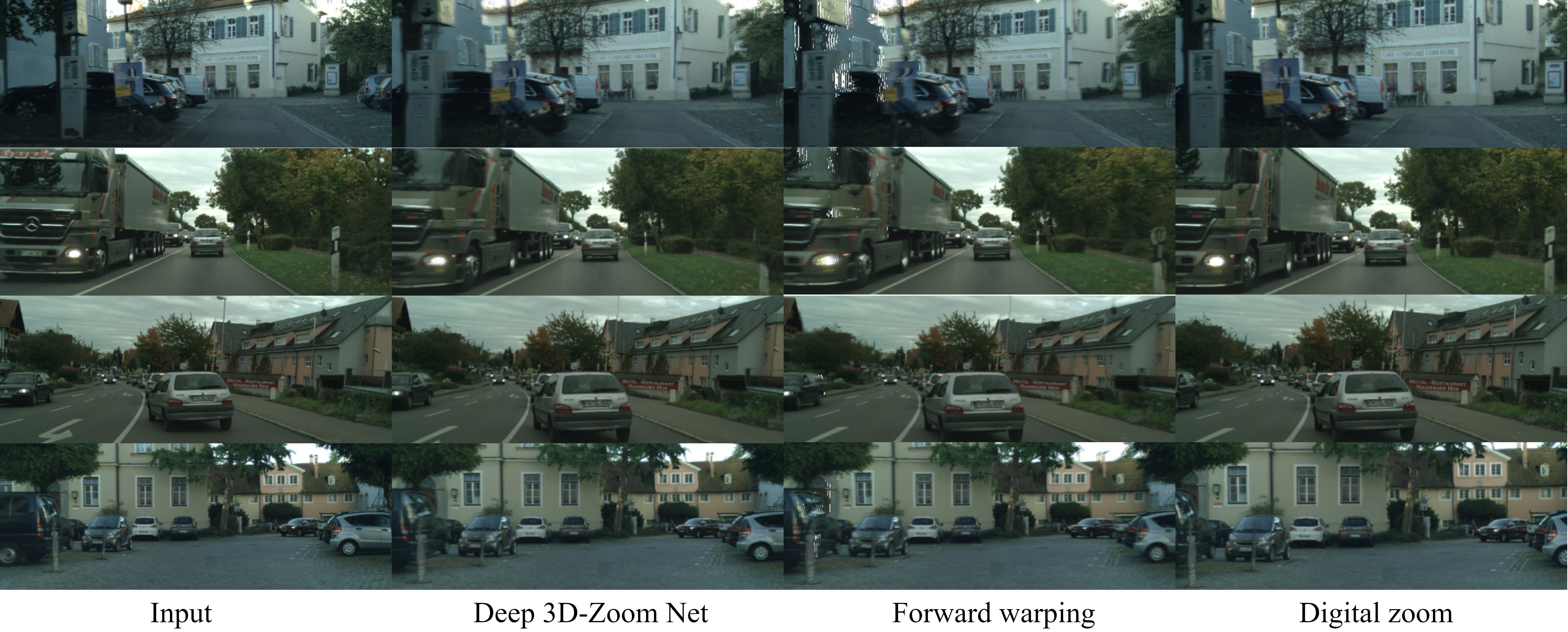}
  \caption{Model performance on Cityscapes dataset. Images generated with different zoom factors showing our network performs well even on unseen scenes. Forward warping, guided by disparity estimation from \cite{gonzalez}, produces blurred, occluded, and deformed results. Digital zoom based on linear interpolation produces uniformly up-scaled images, thus not accounting for 3D geometry.}
  \label{fig:cityscapes}
\end{figure*}

\begin{figure}
  \centering
  \includegraphics[width=0.48\textwidth]{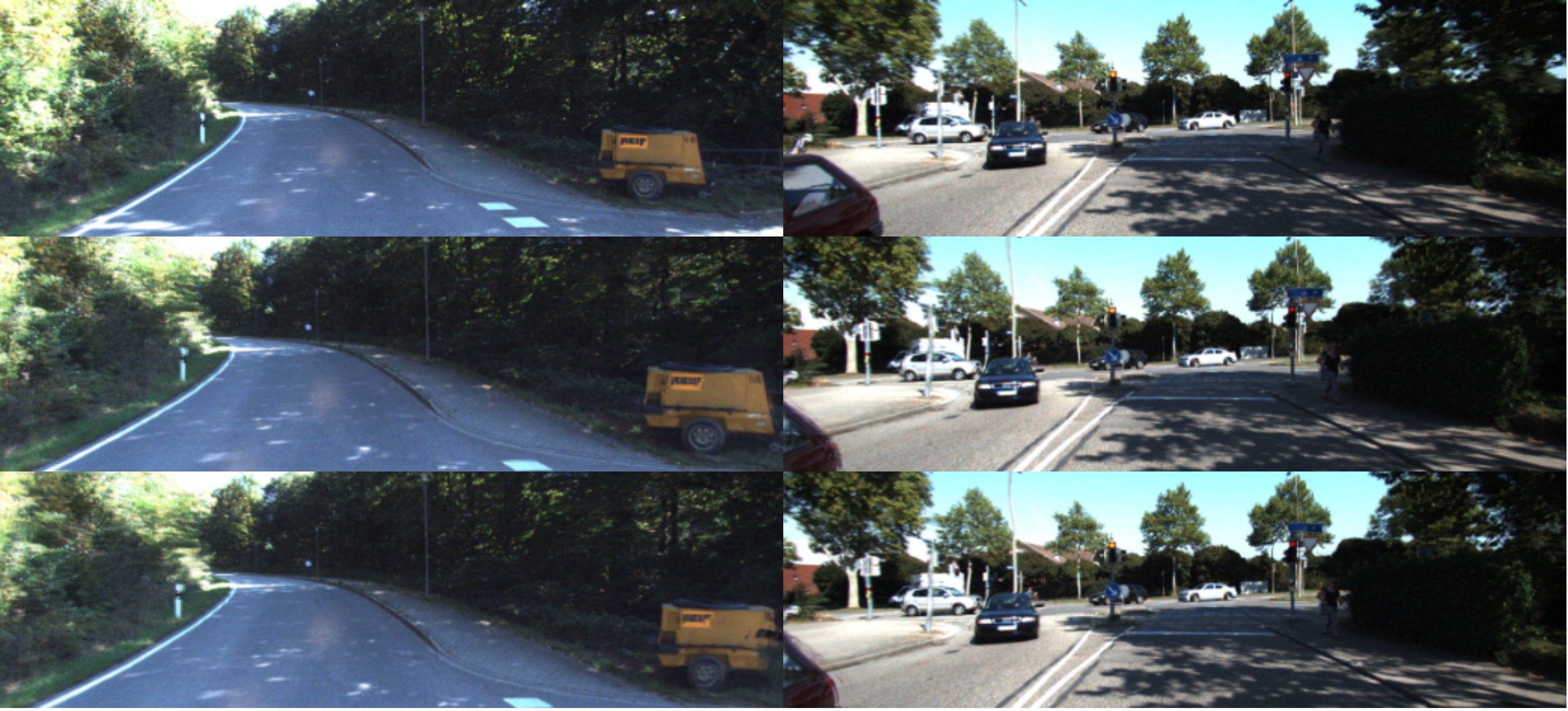}
  \caption{GAN Ablation study. From top to bottom, input images, Deep 3D-Zoom Net with GAN, and Deep 3D-Zoom Net w/o GAN. The adversarial loss helps by reducing ghosting artifacts as can be appreciated in the power generator (right) and car boot (left).}
  \label{fig:ablation1}
\end{figure}

\subsection{KITTI}
We loosely compare our results with the results presented in \cite{geo_aware}, whenever their camera motion was mostly positive in the Z-axis, with the objective of comparing how natural the generated images look. As depicted in Figure \ref{fig:vs_geo} our method generates considerably better natural images, with few or no artifacts. The equivalent zoom factor used in each image generated by our method is 1.6 for the top row, and 2.4 for the bottom row. Our Deep 3D-Zoom Net performs very fast inference on a 1225x370 image in 0.01 seconds on a Titan Xp GPU.

\subsubsection{Ablation studies}
We performed ablation studies to prove that the refinement block, the perceptual loss, and the adversarial loss contribute to improving the final quality of the generated image. As depicted by the qualitative results in Figure \ref{fig:ablation}, each part of our full pipeline improves the overall result. We measure the performance of our networks on the Kitty2015 dataset by using the {\it no-reference} Natural Image Quality Evaluator (NIQE) metric \cite{niqe}. The average values for the 200 frames in the KITTI2015 dataset for 1.5, 2.0 and 2.5 zoom factors are presented in Figure \ref{fig:ablation}, where the lower value is better. As depicted in Figure \ref{fig:ablation}, the most significant change in quality comes with the perceptual loss, as can be seen in the textured areas of the image (e.g. threes and van logos).
Figure \ref{fig:ablation1} shows the benefits of utilizing the adversarial loss. The adversarial loss reduces the ghosting artifacts and extraneous deformations, as they rarely appear in natural images. By utilizing our GAN setting, the mean NIQE score falls from 2.99 to 2.86, demonstrating the effectiveness of the adversarial loss. 

\subsection{Cityscapes}
To prove our model can generalize well to other outdoor datasets, we validate our final model on the challenging Cityscapes dataset. As displayed in Figure \ref{fig:cityscapes} our model shows excellent generalization to the previously unseen data. In addition, we display equivalent results for forward-warping (based on the monocular disparity estimation from \cite{gonzalez}), and digital zoom. Forward warping generates blurred and heavily deformed 3D-zoomed-in images, whereas optical zoom simply does not provide a 3D sensation, as every pixel is up-scaled uniformly. In contrast, our Deep 3D-Zoom Net generates natural-looking 3D-zoomed images.


\section{Conclusions}
We formulated a new image synthesis problem, by constraining it to positive translations in the Z-axis, which we call 3D-zoom, and presented an unsupervised learning solution, called the Deep 3D-Zoom Net. We demonstrated that 3D-zoom can be learned in an unsupervised fashion, by (i) modeling the image synthesis as a blending operation between multiple up-scaled versions of the input image, (ii) by minimizing a novel back re-projection reconstruction loss that facilitates transfer learning from a pre-trained disparity estimation network and accounts for 3D structure and appearance, and (iii) incorporating an adversarial loss to reduce unnaturally synthesized areas. Our Deep 3D-Zoom Net produces naturally looking images for both the KITTI and Cityscapes dataset, establishing a state-of-the-art solution for this class of single image novel view synthesis problem.

{
\bibliographystyle{ieee}
\bibliography{egbib}
}

\end{document}